\documentclass[twocolumn,preprintnumbers,amsmath,amssymb,letter, showpacs, superscriptaddress,pre]{revtex4-1}
\PassOptionsToPackage{caption=false}{subfig}
\usepackage{graphicx}
\usepackage{dcolumn}
\usepackage{bm}
\usepackage{subfig}

\begin{document}

\title{Time Dependence and Density Inversion in Simulations of Vertically Oscillated Granular Layers}

\author{Jonathan Bougie}
 \affiliation{Physics Department,
{Loyola University Chicago, Chicago, IL 60626}}
\author{Veronica Policht}
 \affiliation{Physics Department,
{Loyola University Chicago, Chicago, IL 60626}}
\author{Jennifer Kreft Pearce}
 \affiliation{Chemistry Department,
{University of Texas at Tyler, Tyler, TX 75799}}

\date{\today}

\begin{abstract}

We study a layer of grains atop a plate which oscillates
sinusoidally in the direction of gravity,
using three-dimensional, 
time-dependent numerical solutions
of continuum equations to Navier-Stokes order
as well as hard-sphere molecular dynamics simulations.
For high accelerational amplitudes of the plate, the layer exhibits a steady-state
``density inversion'' in which a high-density portion of the layer
is supported by a lower-density portion.  At low accelerational
amplitudes, the layer exhibits oscillatory time dependence
that is strongly correlated to the motion of the plate.  We show that continuum
simulations yield results consistent with molecular dynamics results
 in both regimes.
\end{abstract}

\pacs{45.70-n,05.65.+b,47.57.Gc}

\maketitle

Although
experimental \cite{rericha, bocquet} and computational 
\cite{ramirez,carrillo2008, bougie2005,bougie2011, lan1995} evidence
demonstrate the potential for hydrodynamic models to describe  important aspects of
granular flow,
a general set of  governing equations for granular media is not yet recognized
\cite{tsimring, eshuis2010}. 
Several proposed rapid granular flow models use binary, inelastic
hard-sphere collision operators in kinetic theory to derive equations of motion
for the continuum fields: number density $n$, velocity ${\bf u}$, and
granular temperature $T$  \cite{goldshtein1,jenkinsandrichman,selaandgoldhirsch}.  
As Eshuis, {\it et al} \cite{eshuis2010} stated in 2010,
``The holy grail question in research on granular dynamics is \cite{jaeger96,goldhirsch2003}, 
To what extent can granular flow be described by a continuum approach?'' 

Density inversion, in which a low-density region near the bottom of a granular layer supports
a higher-density region above it,  has
proven to be significant for the study of granular hydrodynamics.  This phenomenon has been identified in 
vertically shaken layers \cite{lan1995, eshuis2005, brey01, meerson2003, eshuis2010} as well as
in layers flowing parallel to a surface, such as in gravity-driven flow down an incline \cite{taberlet2006,borzsonyi2009}.

In their seminal investigation \cite{lan1995}, Lan and Rosato studied density inversion in vertically vibrated granular media.
A layer of grains with depth $H$ and uniform diameter $\sigma$ 
atop a plate that oscillates sinusoidally
 with frequency $f$ and amplitude $A$ will leave the plate at some time in the oscillation cycle
if the maximum acceleration of the plate $a_{max}=A\left(2\pi f\right)^2$ exceeds 
the acceleration of gravity $g$.  The oscillating plate can be characterized by the
dimensionless parameters $\Gamma=a_{max}/g$ and $f^{*}=f\sqrt{H/g}$.  Lan and Rosato studied density inversion in such a system by conducting soft-sphere discrete element 
method (DEM) simulations and comparing these results to kinetic theory predictions of Richman and Martin \cite{richman1992}.

These continuum predictions did not account for the time dependence of the layer or the plate, but rather
treated the oscillating plate as a source of thermal energy and assumed one-dimensional (1-D) steady-state density and 
temperature distributions as functions of height in the cell.  To characterize the rate of 
kinetic energy input through shaking, they used 
the dimensionless RMS speed of the bottom plate $V_{b}=\left({2\pi f A}\right)/\left({\sqrt{2 \sigma g}}\right)$
as their control parameter.  It has since become common 
to instead use the dimensionless shaking strength \cite{eshuis2005}
\begin{equation} S=2V_b^2=\Gamma A/\sigma=\left(\frac{\Gamma}{f^*}\right)^2\left(\frac{H}{4\pi^2\sigma}\right). \end{equation}

In Fig. 5 of their manuscript, Lan and Rosato chose a fixed ``dimensionless mass hold-up"
$m_t\equiv\frac{N\pi\sigma^2}{6C}=2.5$, where $N$ is the number of particles in the layer, 
and $C$ is the cross-sectional area of the bed.  This corresponds to 
an average layer depth $H\approx4.3\sigma$ as poured.
They investigated shaking strengths
$S=8$ and $S=50$ by varying 
$\Gamma$ and $f^{*}$ in direct proportion, while holding $S$ constant.  For each value of $S$, their DEM simulations exhibited two different regimes of behavior,
depending on $\Gamma$ (or proportionally, $f^{*})$.  

In the high-$\Gamma$ regime, DEM simulations exhibited a nearly steady-state density profile,
in which a high-density region above the plate was supported by a high-temperature but low-density region near the plate. 
Their continuum model produced density and temperature profiles consistent with
this steady-state ``density inversion'' \cite{lan1995}.  
In the low-$\Gamma$ regime, however,
the motion of the layer in DEM simulations exhibited strong time dependence
correlated to the motion of the plate. Similar time dependence at low acceleration is found in conjunction with shocks \cite{bougie2002},
patterns \cite{bizon98}, and compaction \cite{ribiere2005} of granular beds.
Lan and Rosato found that  their 
continuum model could not accurately describe these density profiles \cite{lan1995} in this regime.

Since that time, several studies have focused mainly on the high-$\Gamma$, steady-state regime, providing significant evidence
 that steady-state hydrodynamic solutions can accurately model this regime
\cite{eshuis2005, brey01, meerson2003}.
A further study investigated the transition from a time-independent Leidenfrost
state (in which a high-density crystalline cluster is supported by a low-density, fast-moving
granular gas) to a convection state
using linear stability analysis of
time-dependent continuum equations
\cite{eshuis2010}.  

We believe that the failure of kinetic theory to describe the low-$\Gamma$ regime
was primarily due to the assumption of a steady-state solution.
In our current manuscript, we conduct time-dependent continuum simulations
and compare to molecular dynamics (MD)
simulations.
This comparison is analogous to the procedure used by Lan and Rosato,
with two significant modifications.  First, while Lan and Rosato compared 1-D, 
time-independent kinetic theory predictions to soft-sphere DEM simulations,
we use a 3-D, time-dependent continuum simulation.
We choose a continuum model \cite{jenkinsandrichman} that has previously yielded results
consistent with a hard-sphere MD simulation for shocks \cite{bougie2002} and patterns \cite{bougie2005}
in granular layers, and compare to that same MD simulation. 
Secondly, while Lan and Rosato modeled sinusoidal shaking along all three spatial axes, we model vertical oscillations commonly
found in experiments \cite{eshuis2005, eshuis2010}.  
We demonstrate that continuum simulations including time dependence
are consistent with density profiles from MD simulations for both the high-$\Gamma$
and low-$\Gamma$ regimes.   

We numerically integrate a set of time-dependent continuum equations to
Navier-Stokes order in three spatial dimensions, as proposed by Jenkins and Richman
\cite{jenkinsandrichman} for a dense gas of frictionless (smooth), inelastic hard spheres. 
The granular fluid is contained between impenetrable horizontal plates at the top and bottom of the container.  
The lower plate oscillates sinusoidally between height $z=0$ and $z=2A$, 
and the ceiling is located at a height $L_{z}=320\sigma$ above the 
lower plate.   
Periodic boundary
conditions are used in the horizontal directions $x$ and $y$ 
to eliminate sidewall effects
in a box with horizontal dimensions $L_x = L_y = 22\sigma$.  We use the same grid sizes and numerical techniques as those used in \cite{bougie2005,bougie2011}.

We compare continuum results to results from
a frictionless, inelastic hard-sphere MD simulation.  A version of this
simulation including frictional interactions has been validated 
by comparison to experimentally observed standing wave patterns \cite{bizon98, moon02},
size segregation of binary mixtures \cite{schroter2006}, and fluidization of granular monolayers \cite{gotzendorfer2005}.
The horizontal dimensions of the cell are identical to those in the continuum simulations, with
periodic boundary conditions.  The particles are constrained between a bottom plate which 
oscillates sinusoidally between $z=0$ and $z=2A$, and a ceiling fixed at $z=320\sigma$.

The layer depth as poured $H=4.30\sigma$  
and the normal coefficient of restitution $e=0.90$ are chosen to match
those in \cite{lan1995}.  Each simulation runs for 150 cycles of the plate 
to reach an oscillatory state; number density $n$ 
and granular temperature $T$ are then
calculated at each height in the cell by averaging the following 100 cycles.

\begin{figure}%
\subfloat{{\includegraphics[trim=.55cm 0.9cm 0.9cm .9cm, clip=true,width=.256\textwidth]{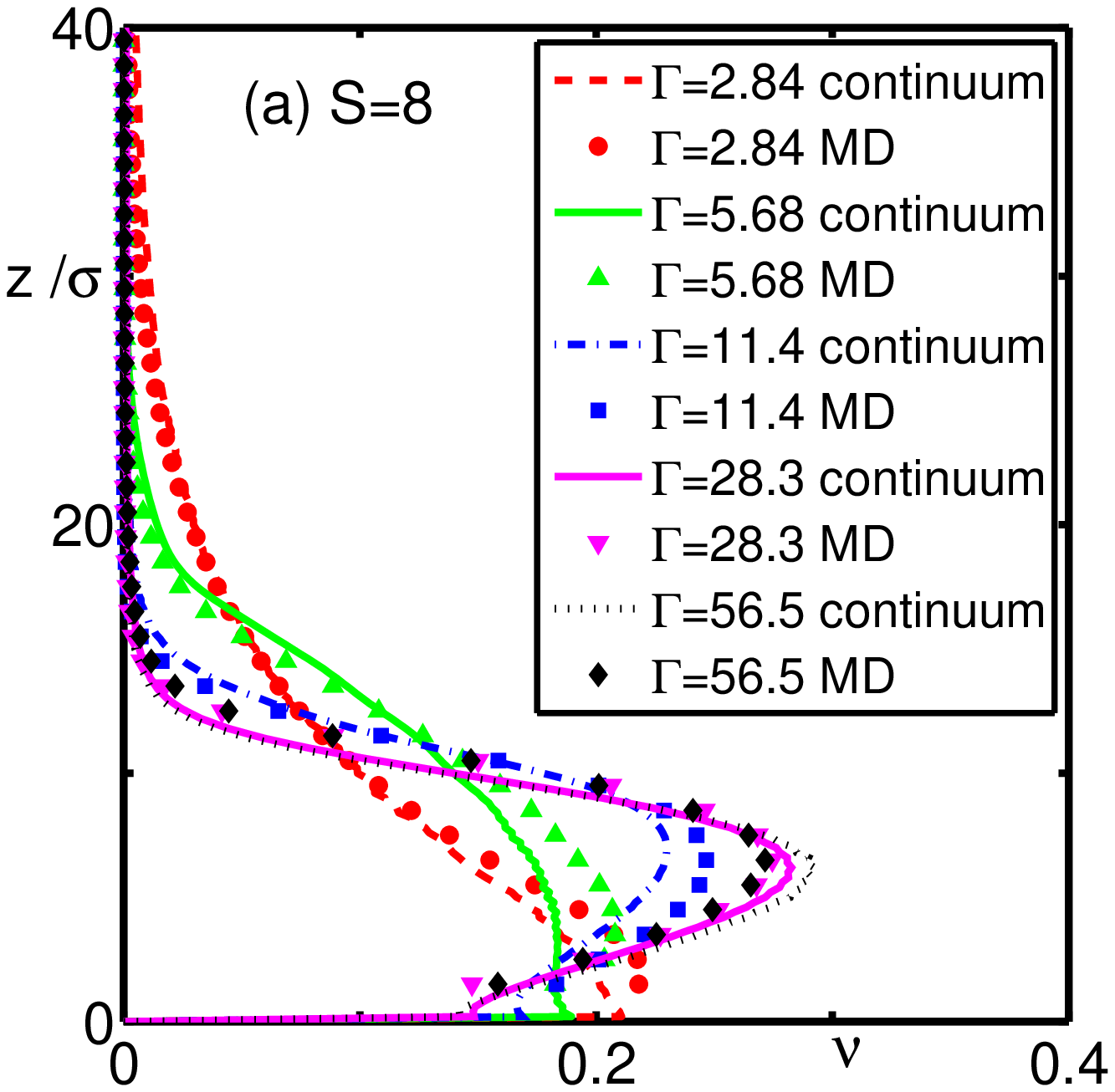}}}
\subfloat{{\includegraphics[trim=1.78cm 0.9cm 0.75cm .9cm, clip=true,width=.235\textwidth]{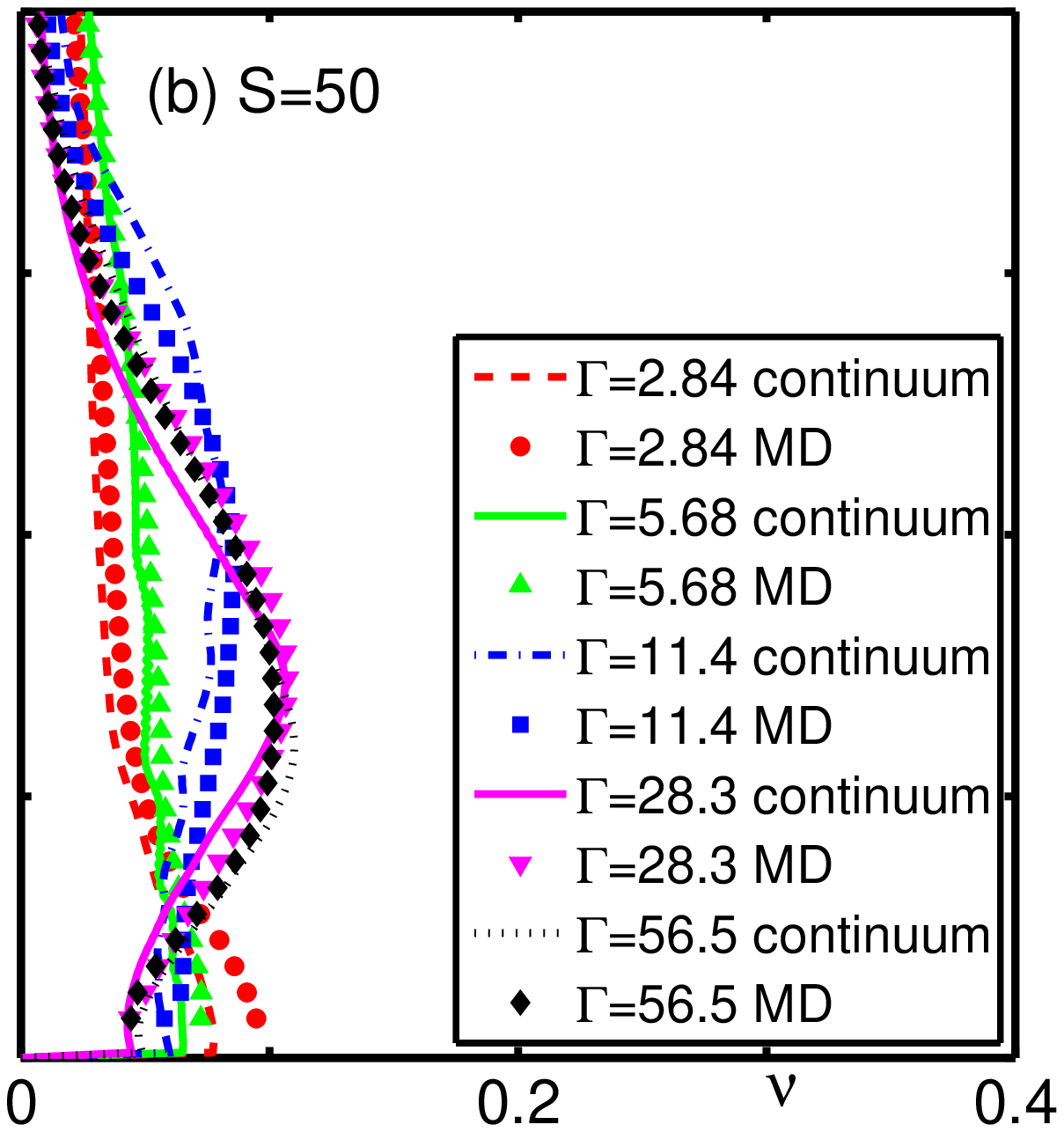}}}\\
\caption{\label{averages} (Color)
Time-averaged volume fraction $\nu$ as a function of height $z/\sigma$ (ordinate) for varying $\Gamma$ corresponding to (a) $S=8$, and (b) $S=50$.  Volume fraction is calculated at eight equal intervals during each oscillation cycle, and these eight times are averaged over 100 cycles of the plate.  Data from MD simulations are represented by symbols, while continuum data are shown as curves.  Although the cell height is $z=320\sigma$, this figure only displays a height up to $z=40\sigma$.  
\vspace{-.25cm}
}
\end{figure}

We use values of $\Gamma$ previously used in \cite{lan1995} to investigate the
time-averaged volume fraction $\nu=\frac{\pi}{6}n\sigma^3$ as a function of height for $S=8$ and $S=50$.  In Fig.~\ref{averages},
both values of $S$ show maximum density near the plate for low accelerational amplitudes $\Gamma=2.84$  and $\Gamma=5.68$,
while higher values of $\Gamma$ show density inversion.  In all cases, 
significant volume fraction is found at at higher values of $z$ for the higher shaking strength $S=50$.
Previous kinetic theory predictions
were consistent with DEM simulations for $\Gamma =28.3$ and $\Gamma=56.5$, but
displayed qualitatively different behavior than DEM simulations for $\Gamma=5.68$ and 
$\Gamma=2.84$ \cite{lan1995}.  By contrast, our time-dependent continuum simulations
show volume fractions similar to those displayed in MD simulations for our entire range of $\Gamma$.

\begin{figure*}%
\hspace{0.0cm}
\subfloat{{\bf $S=8$, $\Gamma=5.68$, $f^{*}=0.66$}} 
\hspace{0.4cm}
\subfloat{{\bf $S=8$, $\Gamma=56.5$, $f^{*}=6.6$}}
\hspace{0.45cm}
\subfloat{{\bf $S=50$, $\Gamma=5.68$, $f^{*}=0.26$}} 
\hspace{0.33cm}
\subfloat{{\bf $S=50$, $\Gamma=56.5$, $f^{*}=2.6$}} \\
\vspace{-.27cm}
\subfloat{{\includegraphics[trim=.18cm .95cm 0.34cm .38cm, clip=true,width=.24\textwidth]{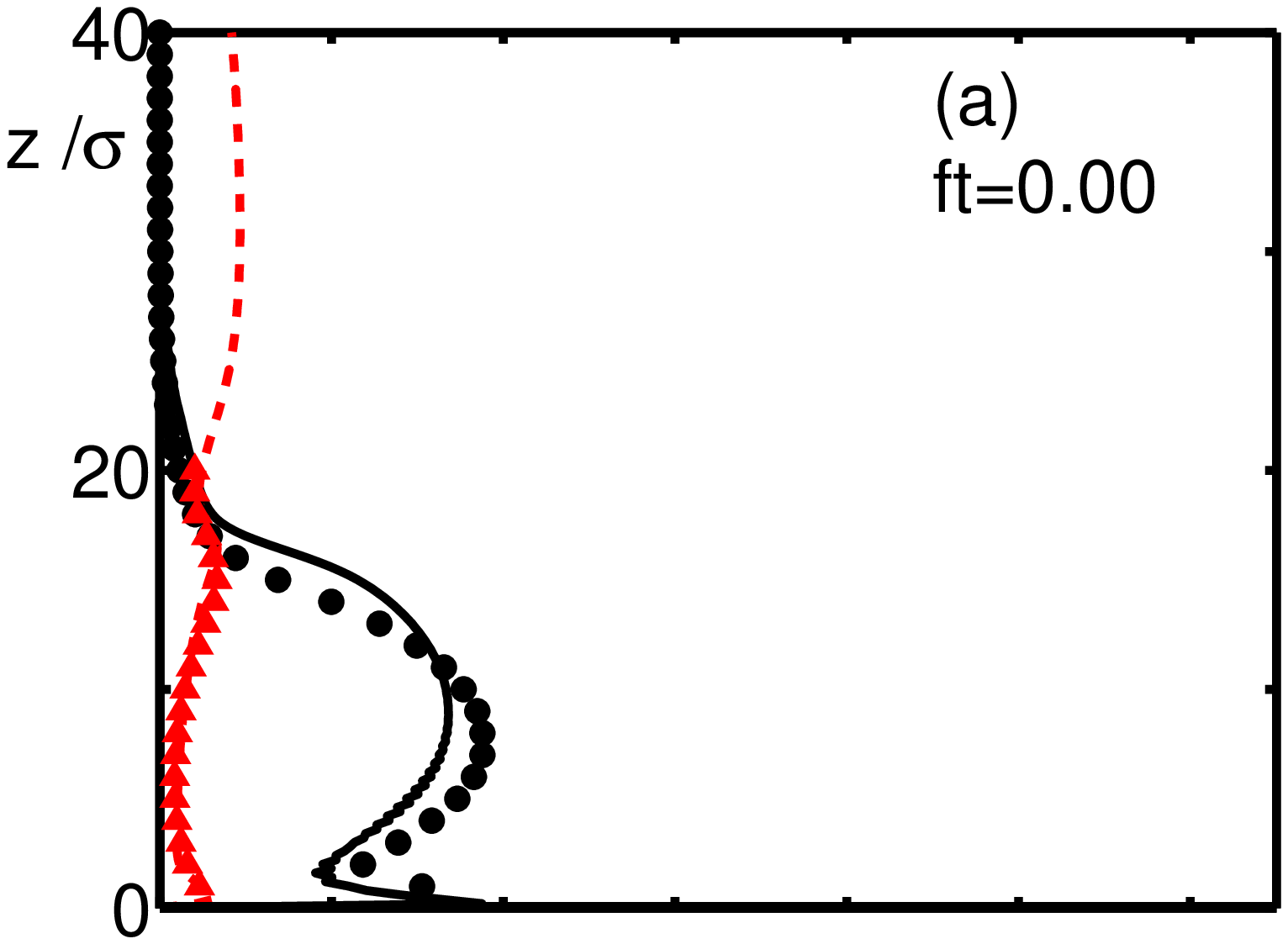}}}
\hspace{.15cm}%
\subfloat{{\includegraphics[trim=.52cm .95cm 0.0cm .38cm, clip=true,width=.24\textwidth]{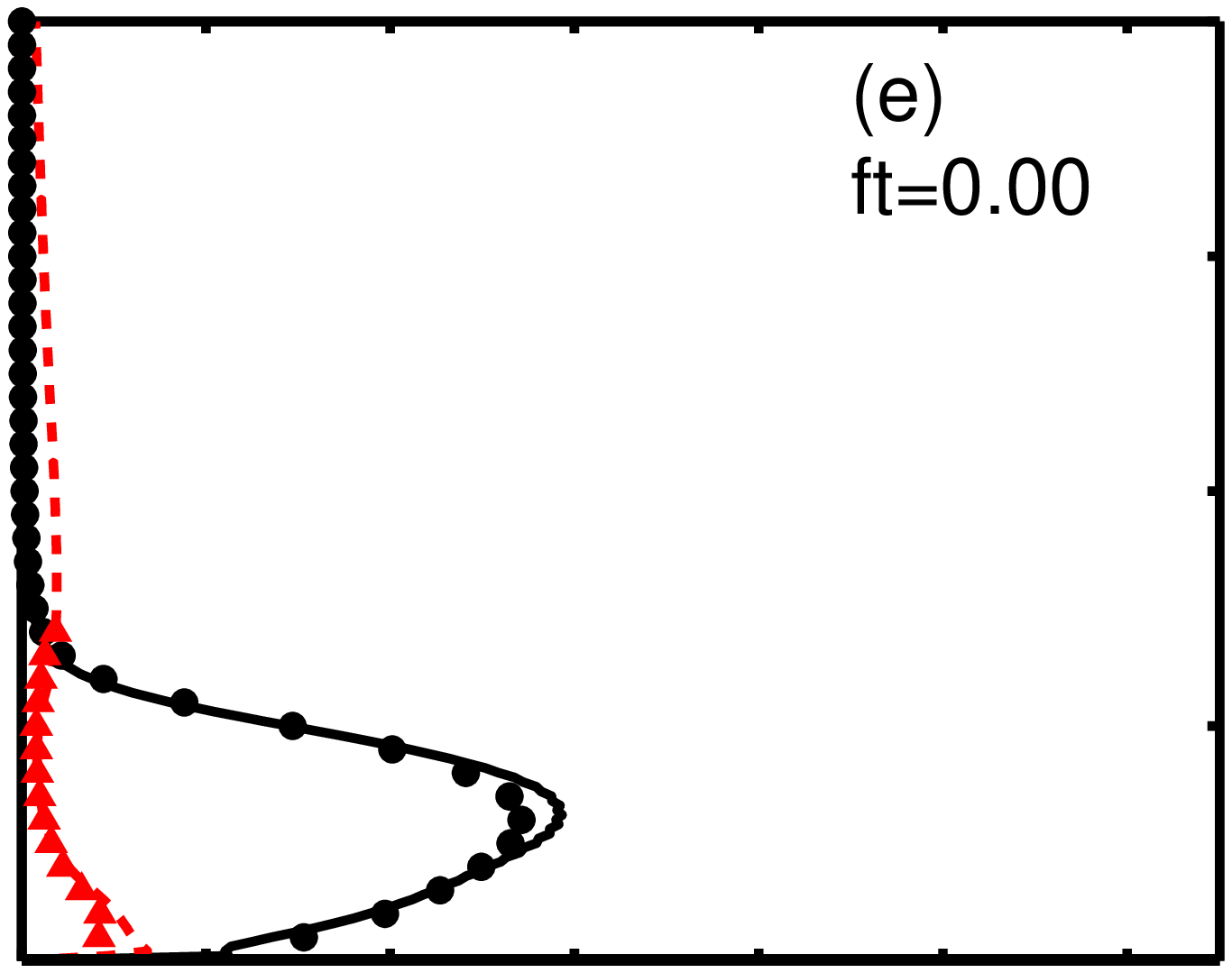}}}
\hspace{.05cm}%
\subfloat{{\includegraphics[trim=.52cm .95cm 0.00cm .38cm, clip=true,width=.24\textwidth]{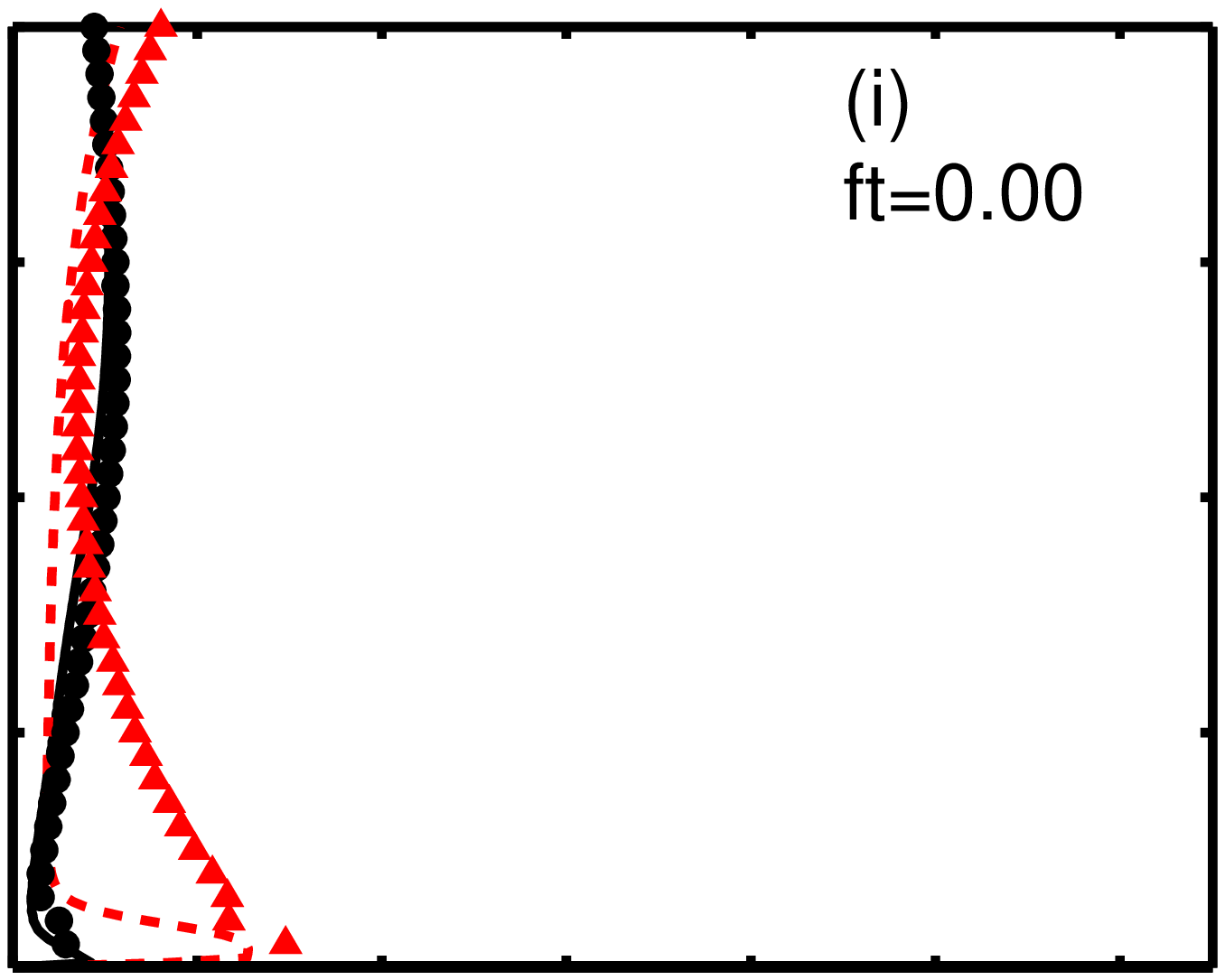}}}
\hspace{.4cm}%
\subfloat{{\includegraphics[trim=.23cm .95cm 0.29cm .38cm, clip=true,width=.24\textwidth]{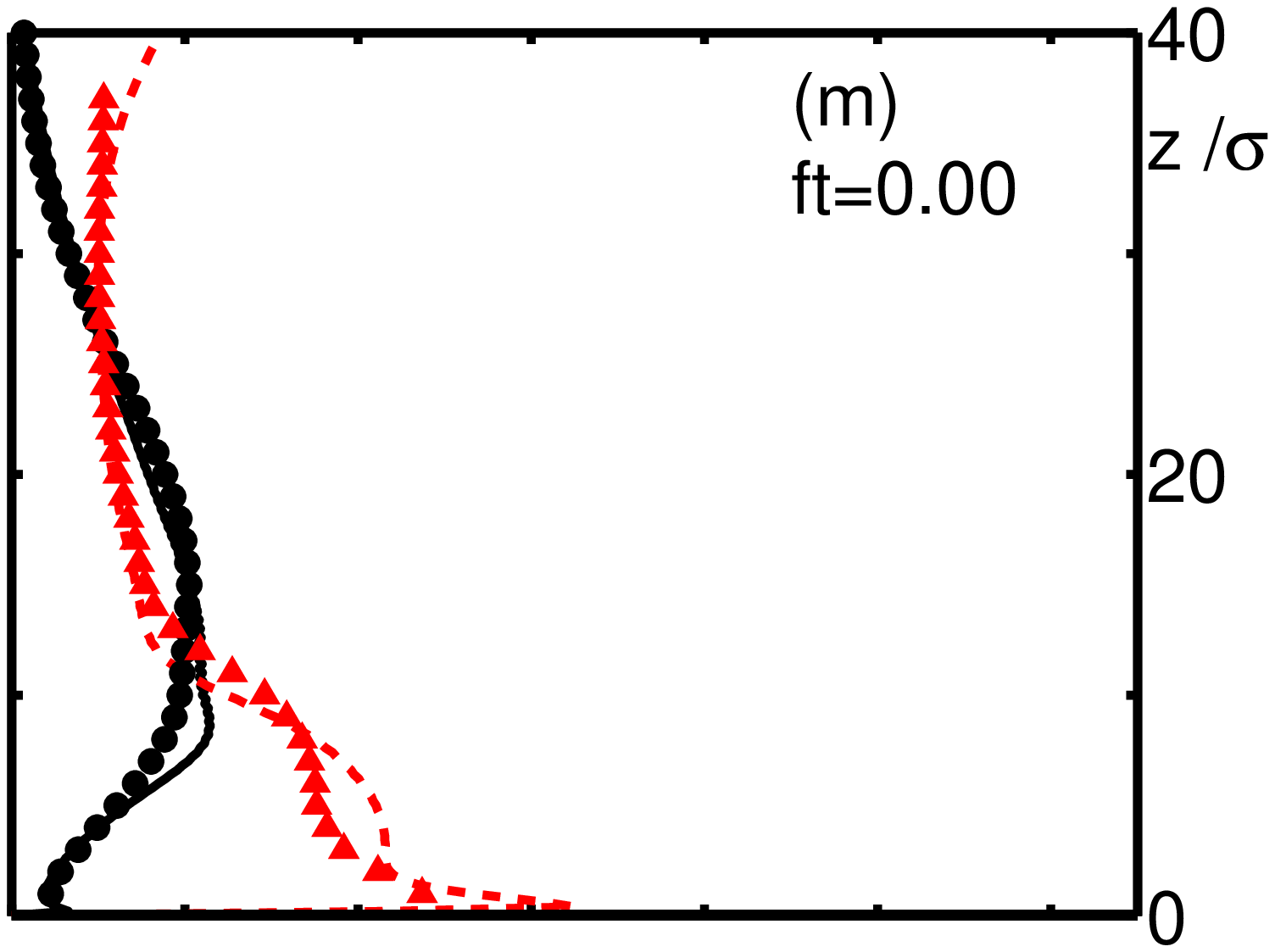}}}\\
\vspace{-.4cm}%
\subfloat{{\includegraphics[trim=.18cm .95cm 0.34cm .38cm, clip=true,width=.24\textwidth]{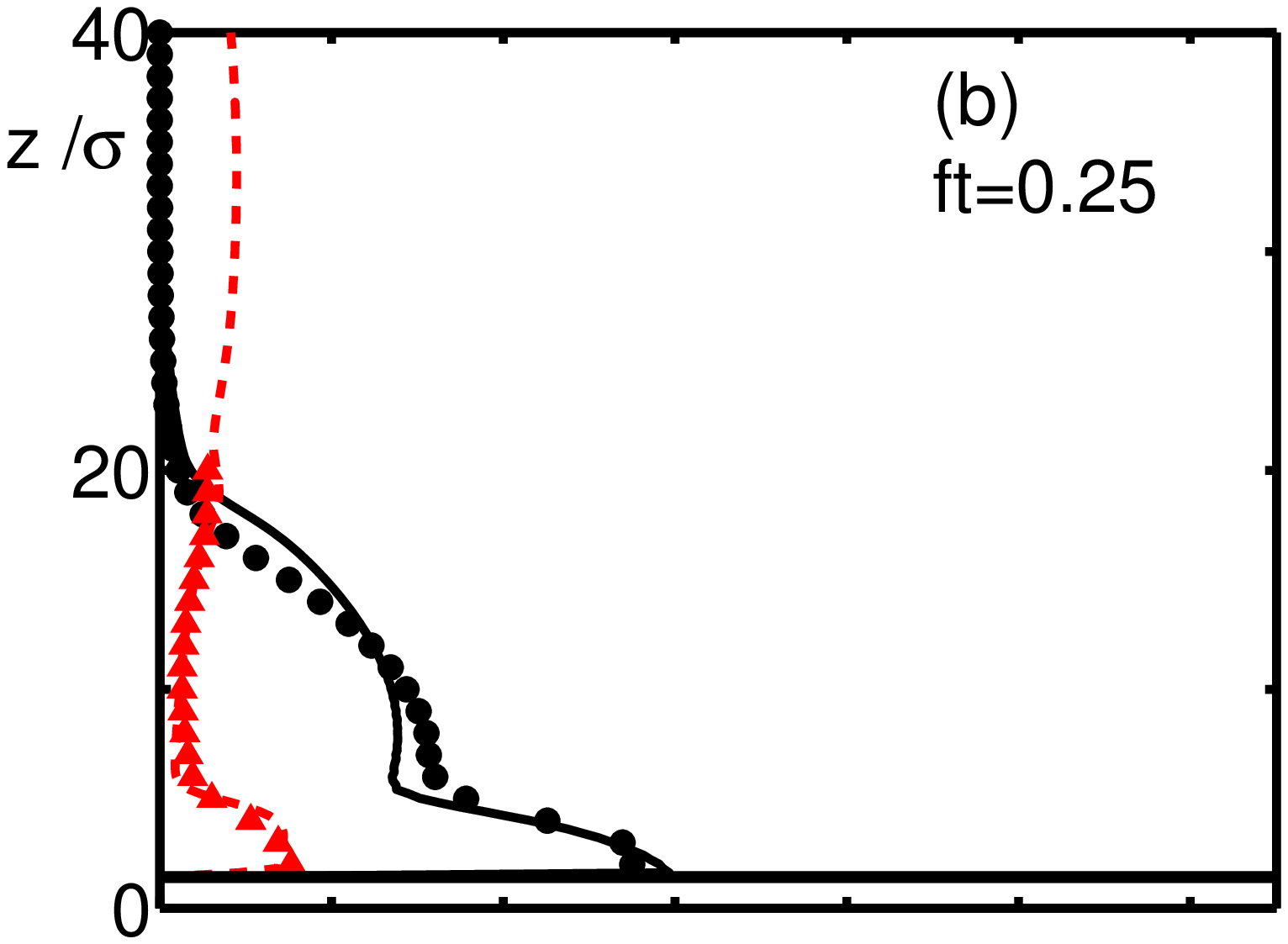}}}
\hspace{.15cm}%
\subfloat{{\includegraphics[trim=.52cm .95cm 0.0cm .38cm, clip=true,width=.24\textwidth]{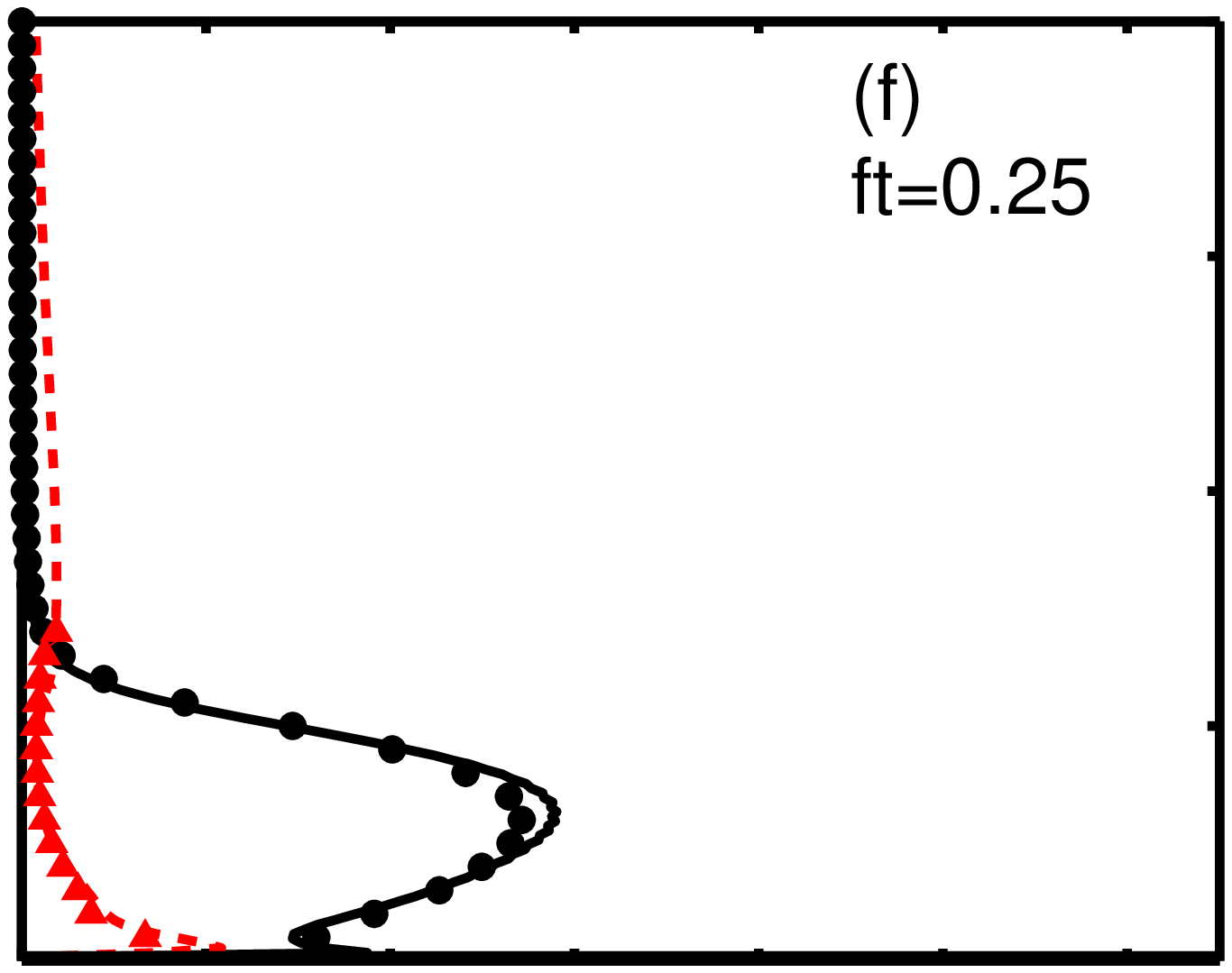}}}
\hspace{.05cm}%
\subfloat{{\includegraphics[trim=.52cm .95cm 0.0cm .38cm, clip=true,width=.24\textwidth]{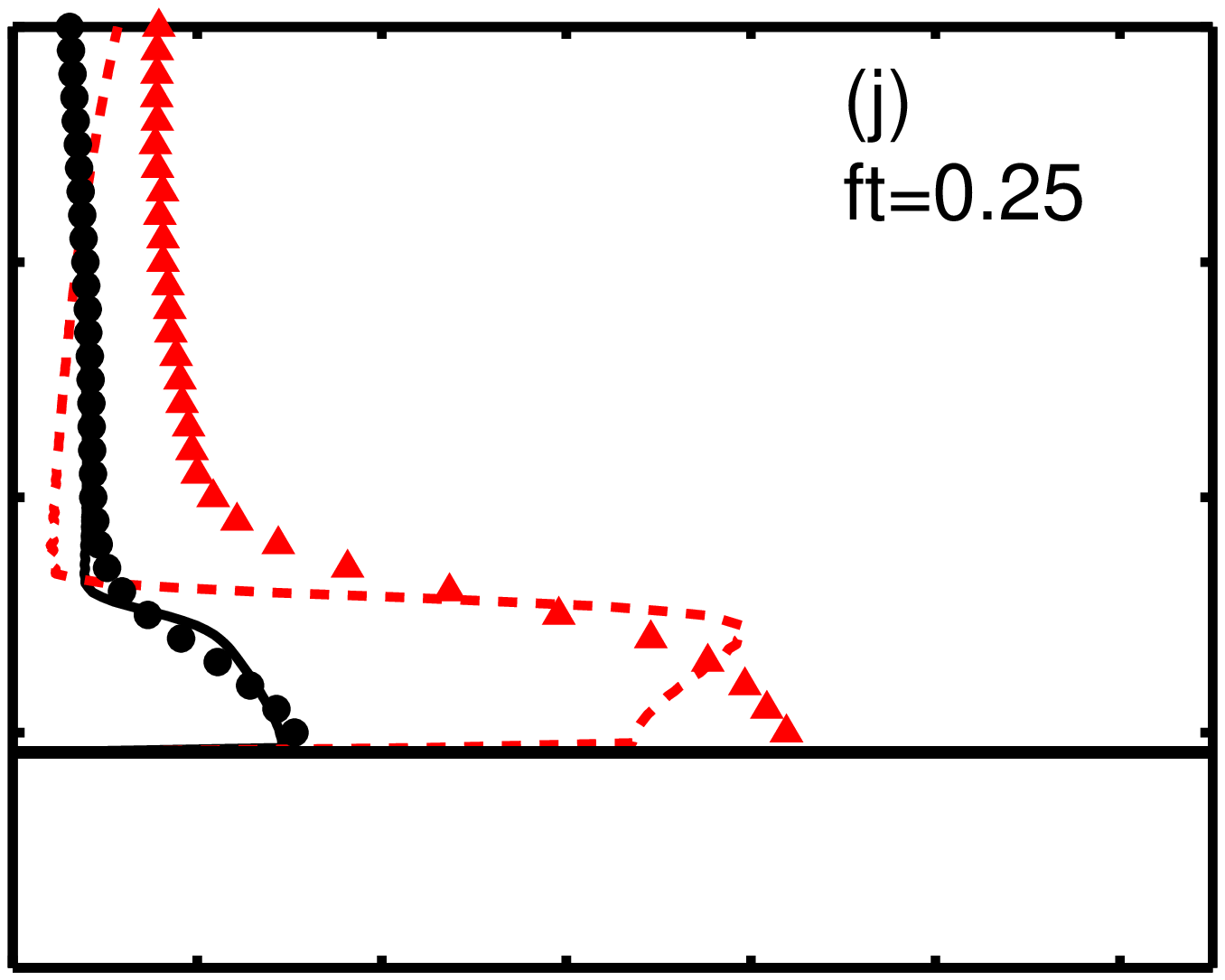}}}
\hspace{.4cm}%
\subfloat{{\includegraphics[trim=.23cm .95cm 0.29cm .38cm, clip=true,width=.24\textwidth]{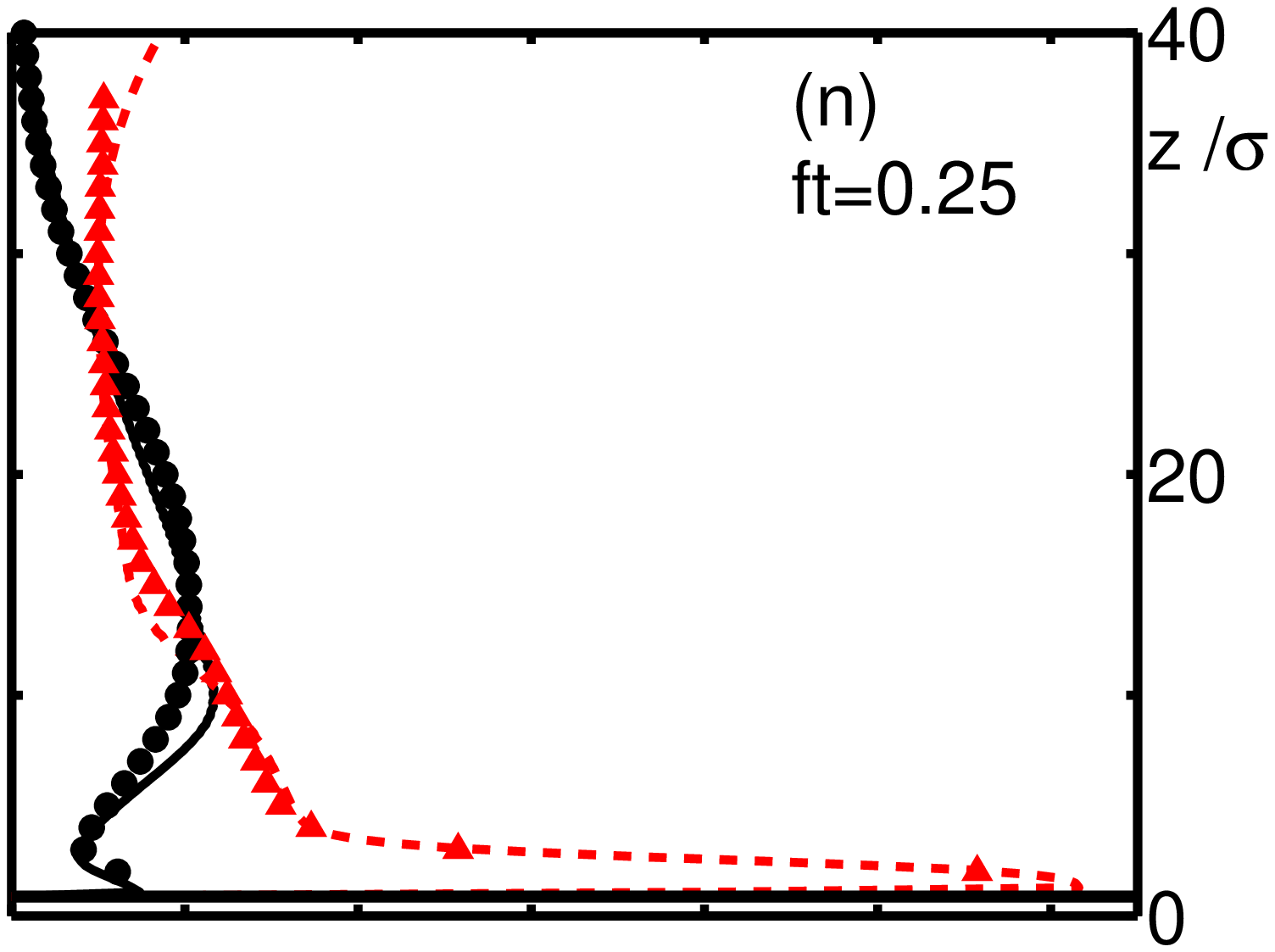}}}\\
\vspace{-.4cm}%
\subfloat{{\includegraphics[trim=.18cm .95cm 0.34cm .38cm, clip=true,width=.24\textwidth]{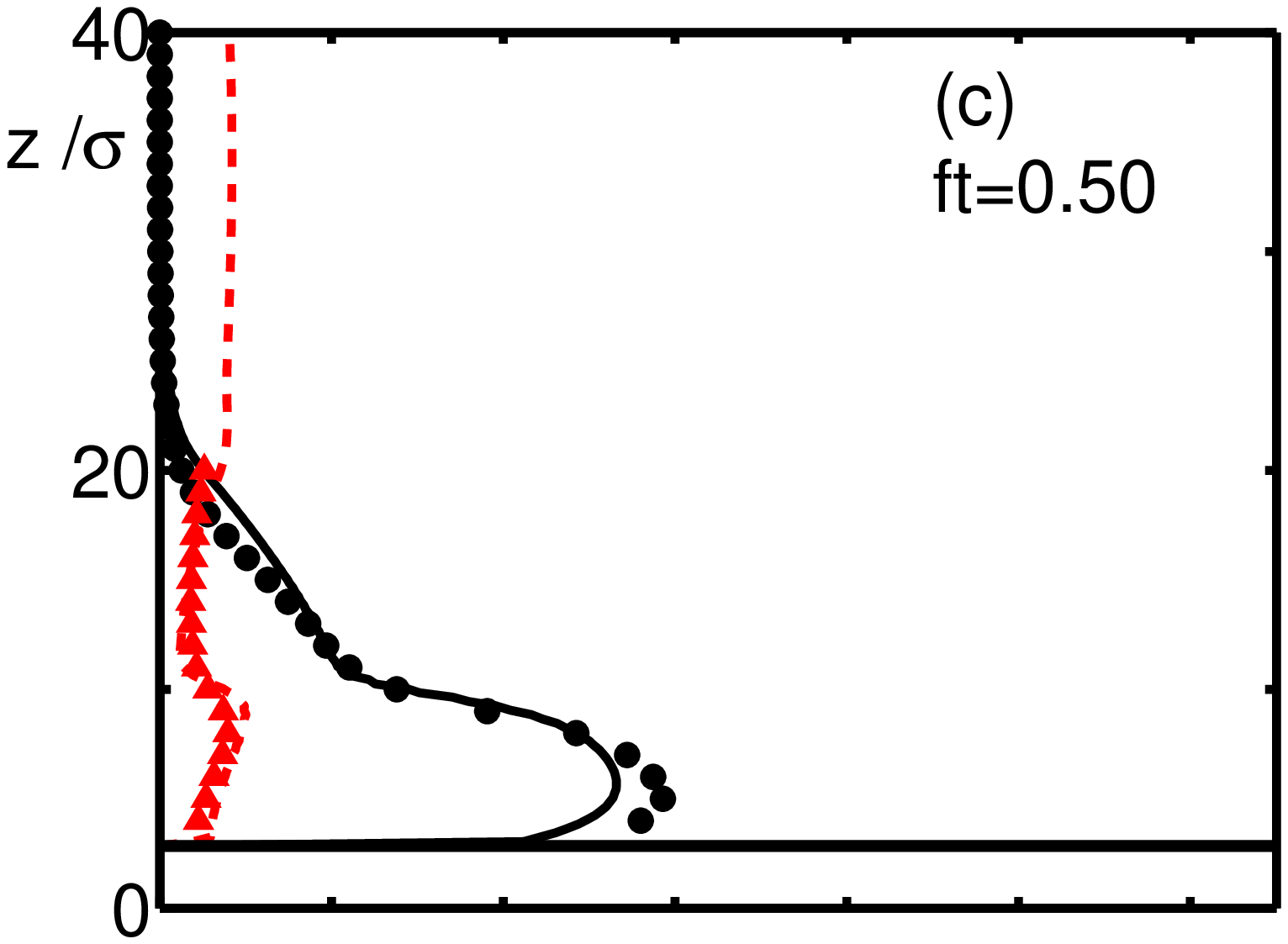}}}
\hspace{.15cm}%
\subfloat{{\includegraphics[trim=.52cm .95cm 0.0cm .38cm, clip=true,width=.24\textwidth]{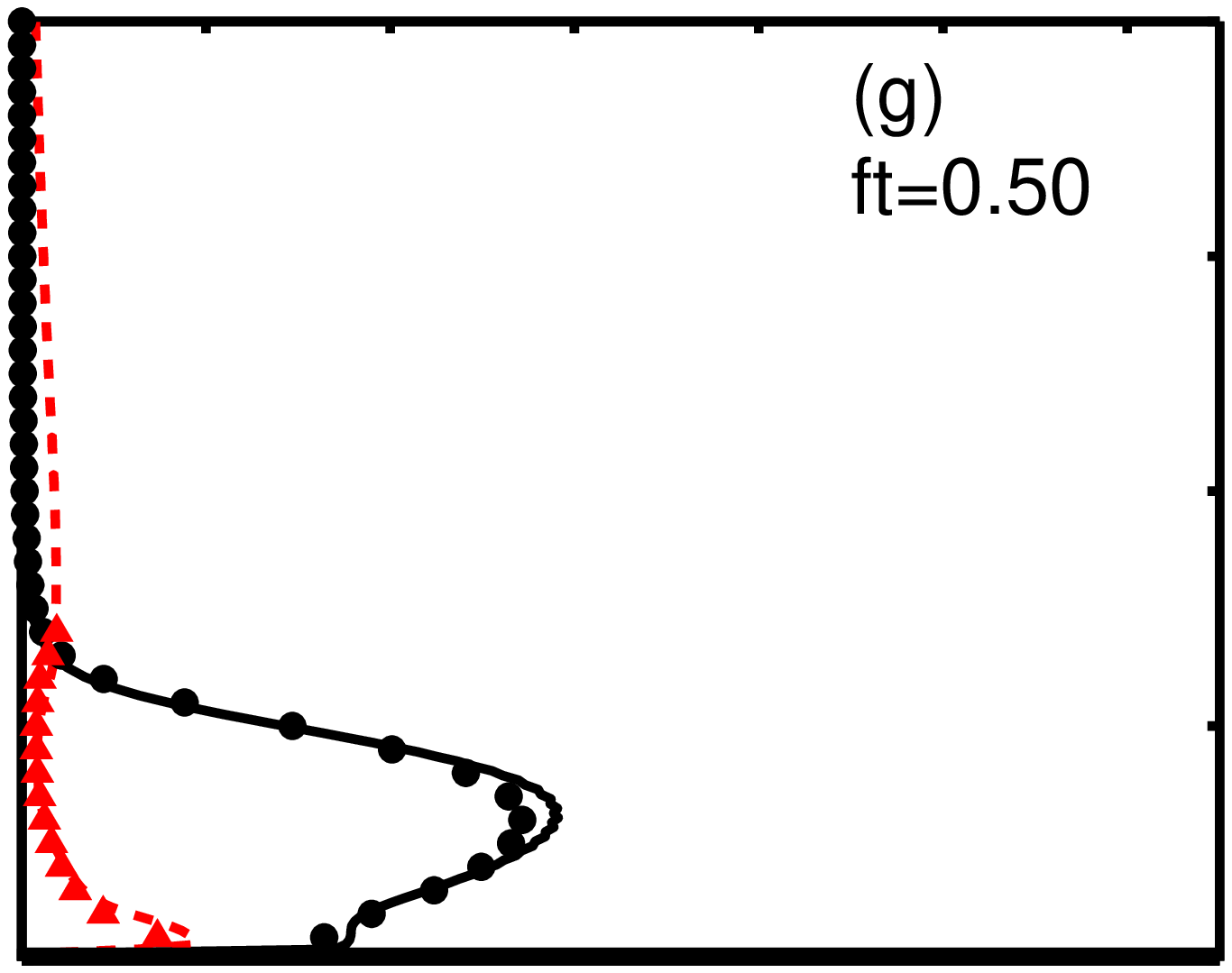}}}
\hspace{.05cm}%
\subfloat{{\includegraphics[trim=.52cm .95cm 0.00cm .38cm, clip=true,width=.24\textwidth]{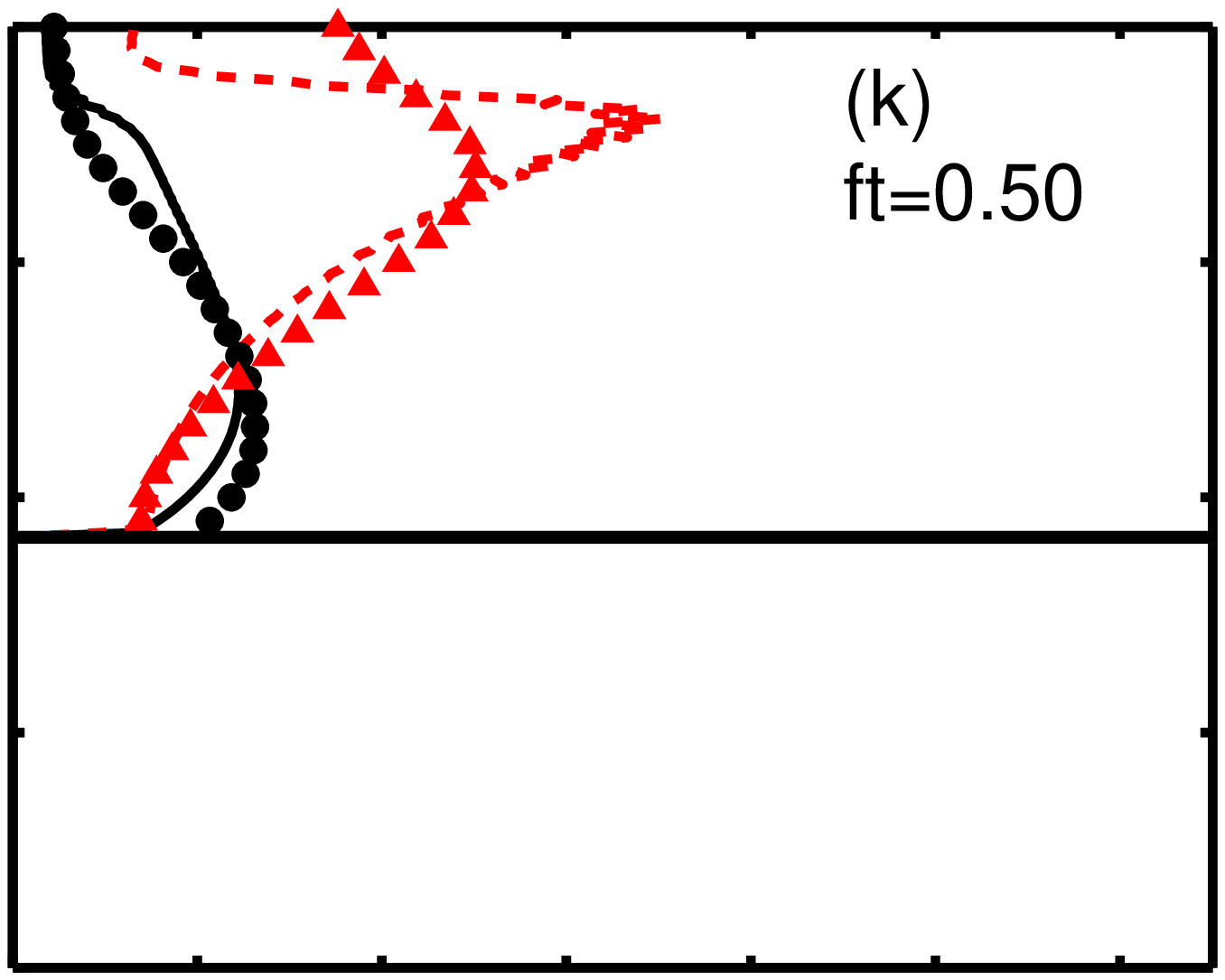}}}
\hspace{.4cm}%
\subfloat{{\includegraphics[trim=.23cm .95cm 0.29cm .38cm, clip=true,width=.24\textwidth]{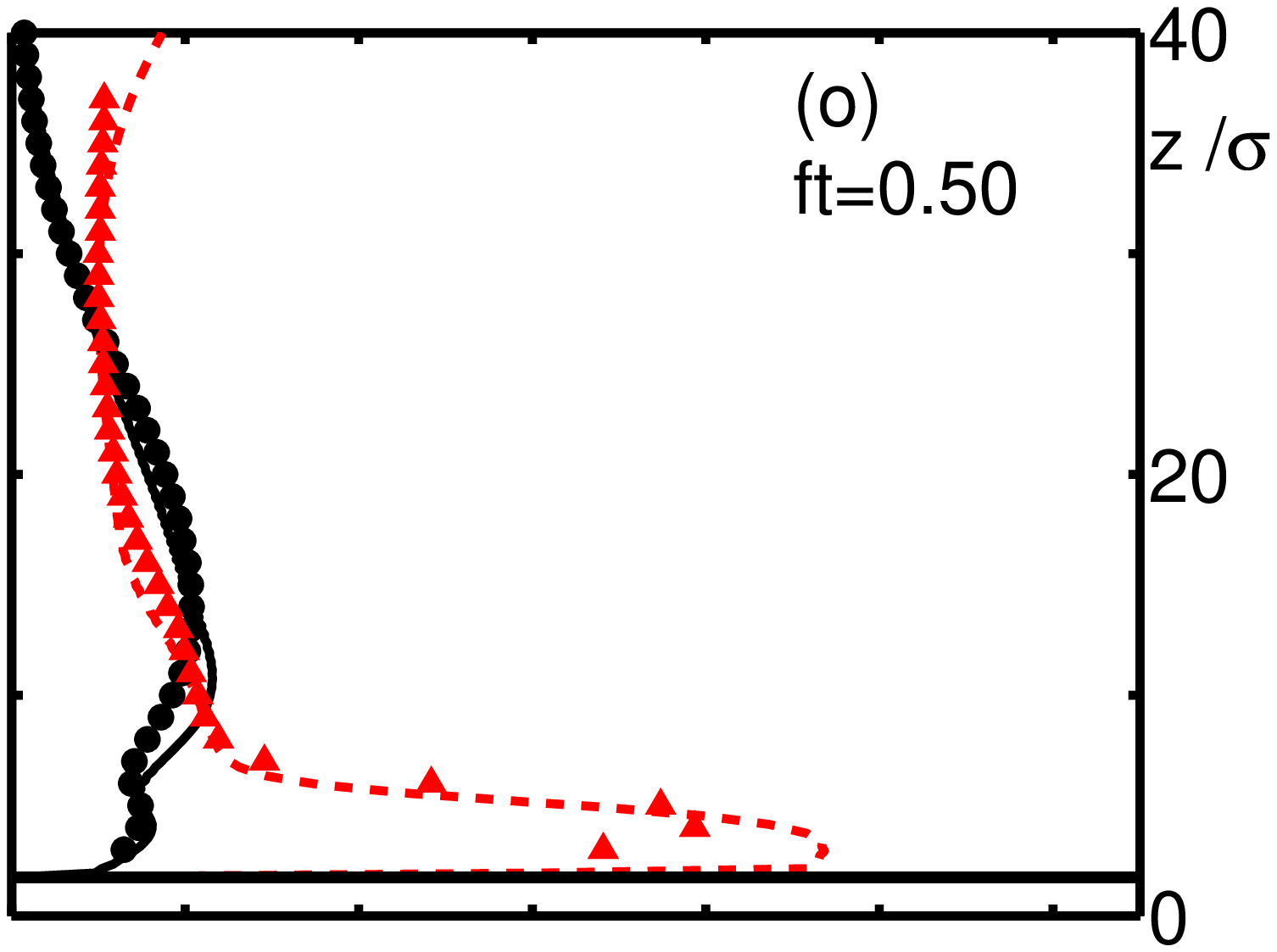}}}\\
\vspace{-.4cm}%
\subfloat{{\includegraphics[trim=.18cm .08cm 0.34cm .38cm, clip=true,width=.24\textwidth]{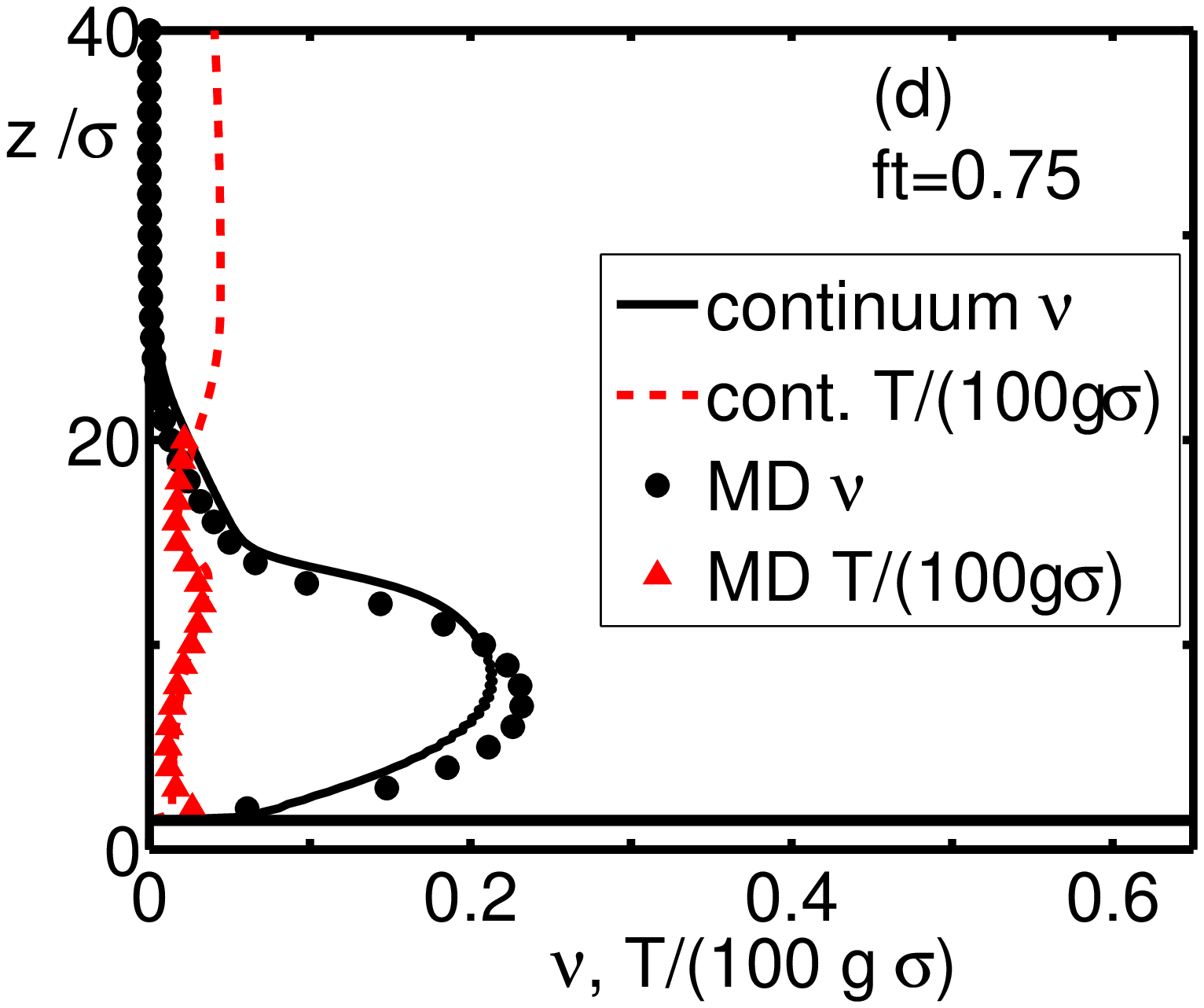}}}
\hspace{.15cm}%
\subfloat{{\includegraphics[trim=.52cm .08cm 0.0cm .38cm, clip=true,width=.24\textwidth]{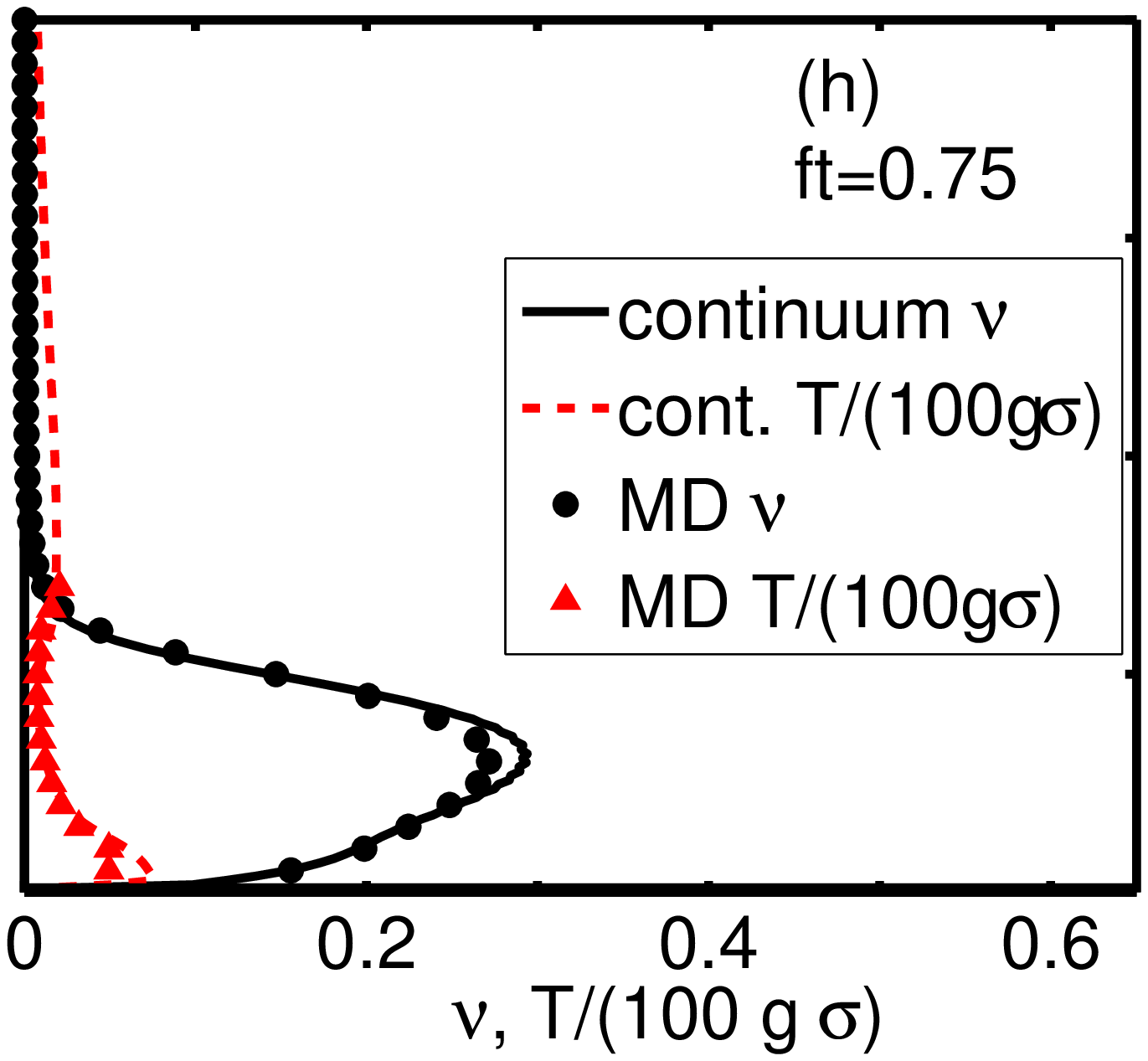}}}
\hspace{.05cm}%
\subfloat{{\includegraphics[trim=.52cm .08cm 0.0cm .38cm, clip=true,width=.24\textwidth]{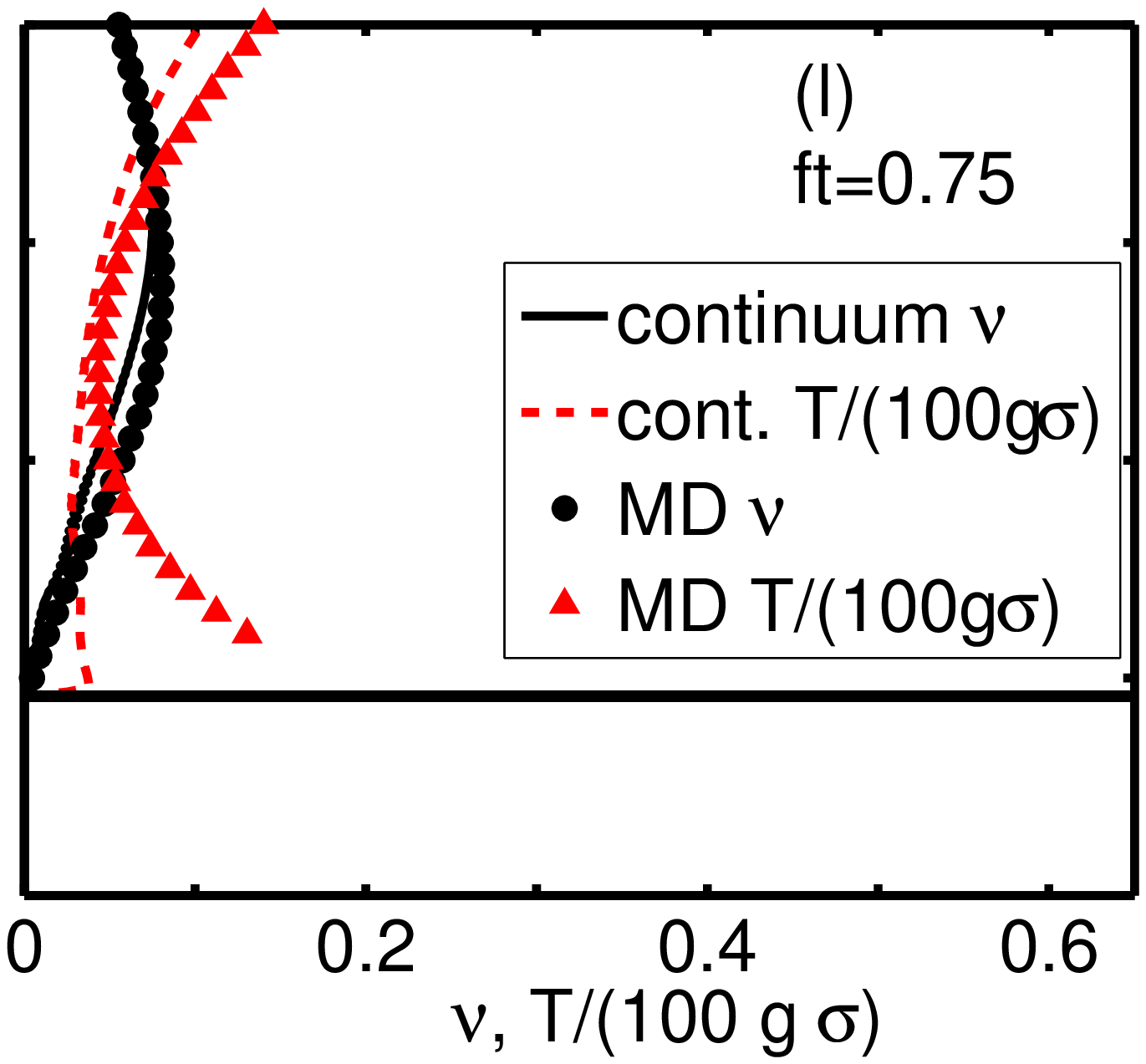}}}
\hspace{.4cm}%
\subfloat{{\includegraphics[trim=.23cm .08cm 0.29cm .38cm, clip=true,width=.24\textwidth]{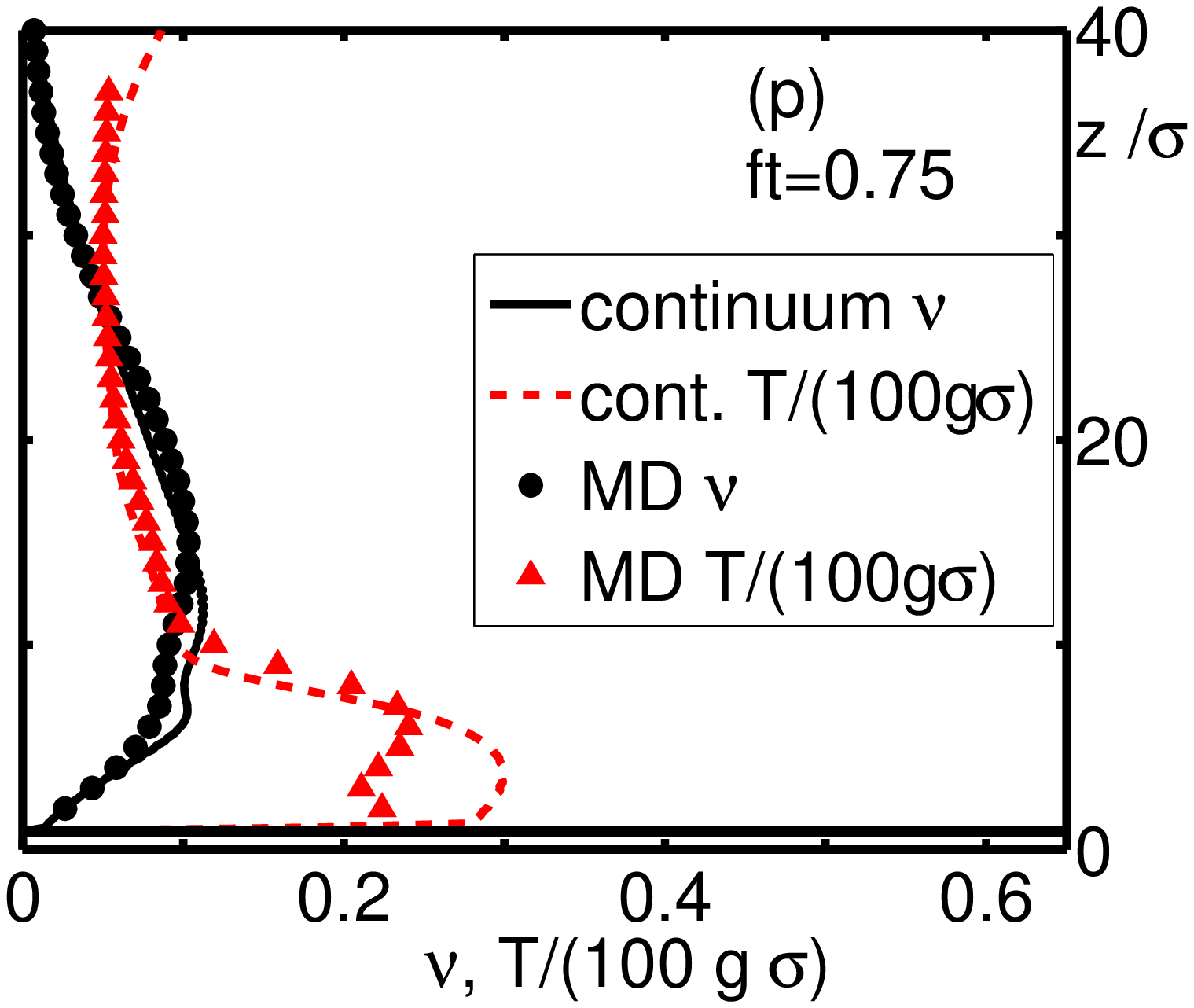}}}
\vspace{-0.11cm}%
\caption{\label{snapshots} 
(Color Online) Volume fraction $\nu$ and dimensionless granular temperature $T/\left(100 g \sigma\right)$ as functions of dimensionless height $z/\sigma$ (ordinate) at four times $ft$ 
in the oscillation cycle.  The first two columns correspond to $S=8$, with the
left-most column (a-d) representing the ``low-$\Gamma$ regime'' ($\Gamma=5.68$, $f^{*}=0.66$) 
and the second column (e-h) representing the ``high-$\Gamma$ regime'' ($\Gamma=56.5$, $f^{*}=6.6$)\cite{lan1995}.  
The final two columns correspond to $S=50$, with the low-$\Gamma$ regime ($\Gamma=5.68$, $f^{*}=0.26$) in subplots (i-l) and the high-$\Gamma$
  regime ($\Gamma=56.5$, $f^{*}=2.6$) in subplots (m-p).
Curves indicate data from continuum simulations, points indicate MD data, and the oscillating bottom of the cell is shown as a solid horizontal black line.  Due to the need for 
a sufficient number of particles for statistical calculations, $T$ in MD simulations is only calculated when $\nu \geq 0.01$.
\vspace{-.1cm}
}
\end{figure*}

We examine the time dependence of two sets of parameters for $S=8$
in Fig~\ref{snapshots} a-h. 
One set ($\Gamma=56.5$, $f^{*}=6.6$)
corresponds to what Lan and Rosato called the ``high $\Gamma$ regime'', while the
other ($\Gamma=5.68$, $f^{*}=0.66$) falls within the ``low $\Gamma$ regime''.

For the low-$\Gamma$ regime (Fig~\ref{snapshots} a-d), the volume fraction and temperature profiles
show significant oscillatory time dependence, correlated to the plate motion
The high-$\Gamma$ regime  (Fig~\ref{snapshots} e-h) shows nearly steady-state profiles, 
in which a higher-temperature but lower-density region near the plate supports a cooler, higher-density region 
away from the plate ($z>5\sigma$).
Continuum simulations produce volume fraction and temperature profiles consistent with MD simulations in both regimes.

Similarly, we examine $\nu$ and $T$ as functions of height for a low-$\Gamma$ case ($\Gamma=5.68$, $f^{*}=0.26$)
and a high-$\Gamma$ case ($\Gamma=56.5$, $f^{*}=2.6$)
corresponding to $S=50$ in Fig.~\ref{snapshots} i-p.  
Interestingly, the temperature profile is strongly time-dependent in both cases.  However, while 
the low-$\Gamma$ case again shows significant oscillatory time dependence in volume fraction,
the high-$\Gamma$ volume fraction again shows a nearly steady state density-inverted profile.
There is a larger discrepancy between temperature predictions from continuum and MD for the low-$\Gamma$ case in which
the temperature is strongly anisotropic \cite{lan1995}.  However, even in this case, the scalar temperature in the continuum
simulation is comparable to that found in MD simulations, and volume fraction profiles are nearly identical in both cases.

\begin{figure}[h]%
{\includegraphics[trim=.0cm 0.2cm 1.55cm 0.4cm, clip=true,width=.26\textwidth]{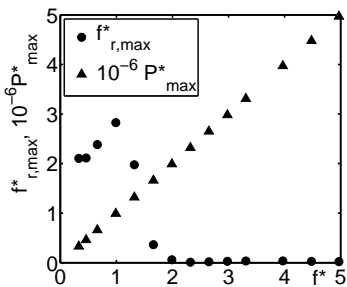}}
\caption{\label{power}
Dimensionless maximum $P^{*}_{max}$ of the power spectrum (circles) and the dimensionless 
dominant
 frequency $f^{*}_{r,max}$ at which this power is found (triangles) as functions of 
dimensionless driving frequency 
$f^*$ in continuum simulations.  $P^{*}_{max}$ is reduced by a factor of $10^6$ to fit on the same scale as $f^{*}_{r,max}$.
}
\end{figure}

These results indicate that time-dependent continuum simulations can capture a transition from
an oscillatory, low-$\Gamma$ regime to a steady-state density inversion at high $\Gamma$.  
This transition was absent in previous time-independent continuum simulations \cite{lan1995}.
To examine the nature of this transition in our continuum simulations, we 
vary $\Gamma$ for constant $S=8$, and measure the height of the maximum volume fraction $z_{max}$ as a function
of time over 100 cycles of the plate, sampled eight times per cycle.  Taking a discrete-time Fourier transform of this time sequence allows us to
calculate a power spectrum as a function of response frequency $f_r$ within the range $f/100\leq f_r \leq 4f$, where $f$ is the driving frequency.  
The dominant response frequency $f_{r,max}$ is the frequency with the maximum power $P_{max}$.   We calculate the 
maximum dimensionless power $P_{max}^{*}=P_{max}/g\sigma$ and corresponding dimensionless response frequency $f^{*}_{r,max}=f_{r,max}\sqrt{H/g}$, 
and plot them as functions of $f^{*}$ in Fig.~\ref{power}.
$P_{max}^{*}$ is relatively high for low driving frequencies, indicating significant time dependence at
$f^{*}_{r,max}$.  However, $P_{max}^{*}$ then drops to near zero 
for $f^{*}\gtrsim2$, indicating a transition to a nearly time-independent state.  The dominant dimensionless
response frequency lies along the line $f^{*}_{r,max}=f^{*}$ both below and above the transition, suggesting
 that the frequency response of the layer is identical with the driving frequency through the transition.  The
transition occurs when the driving frequency exceeds a critical value, at which point, the amplitude of response drops significantly.  

\begin{figure}[h]%
\subfloat{{\includegraphics[trim=.0cm 0.cm 1.55cm 0.2cm, clip=true,width=.245\textwidth]{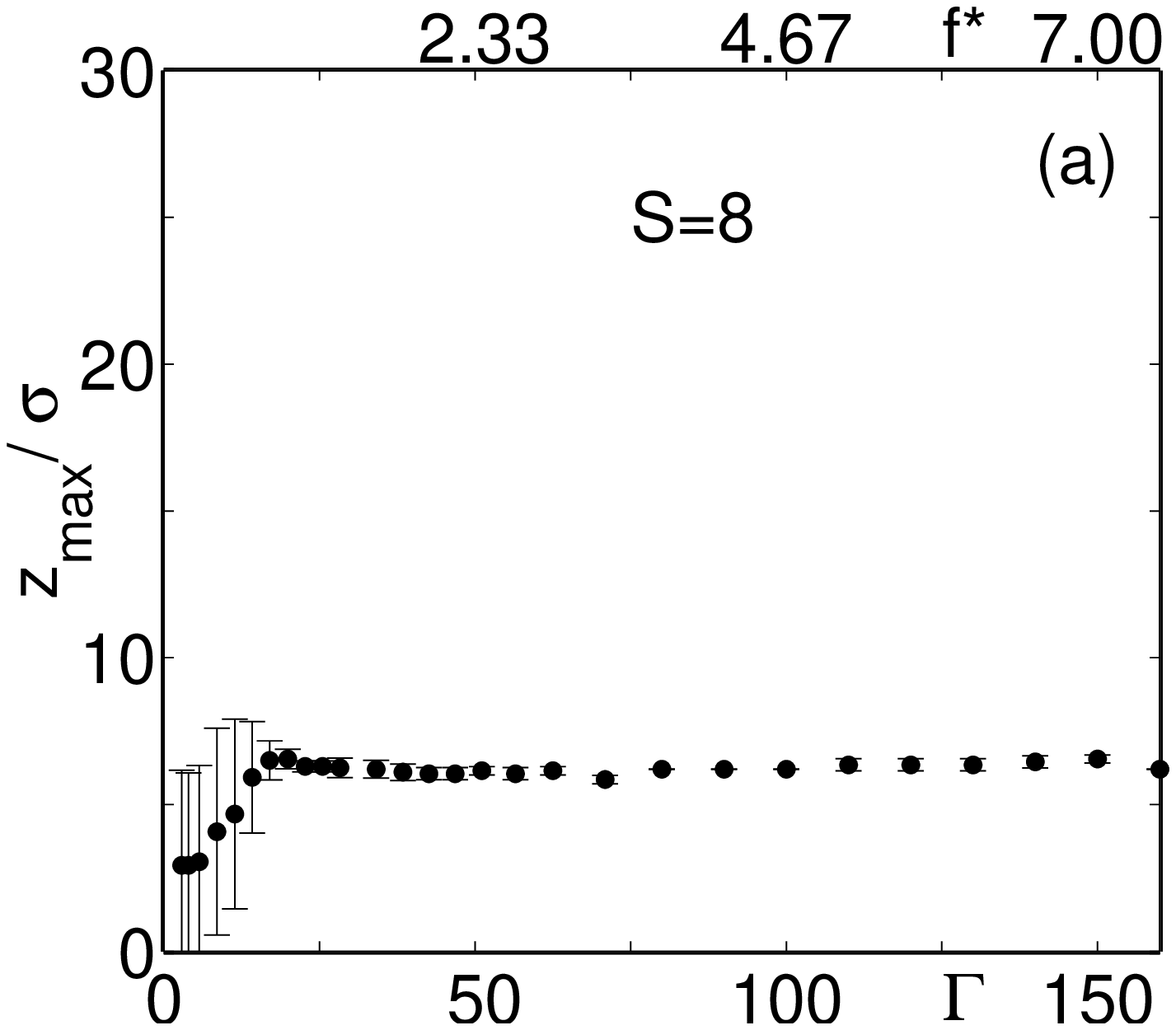}}}
\hspace{-0.2cm}%
\subfloat{{\includegraphics[trim=1.55cm 0.cm 0.0cm 0.2cm, clip=true,width=.245\textwidth]{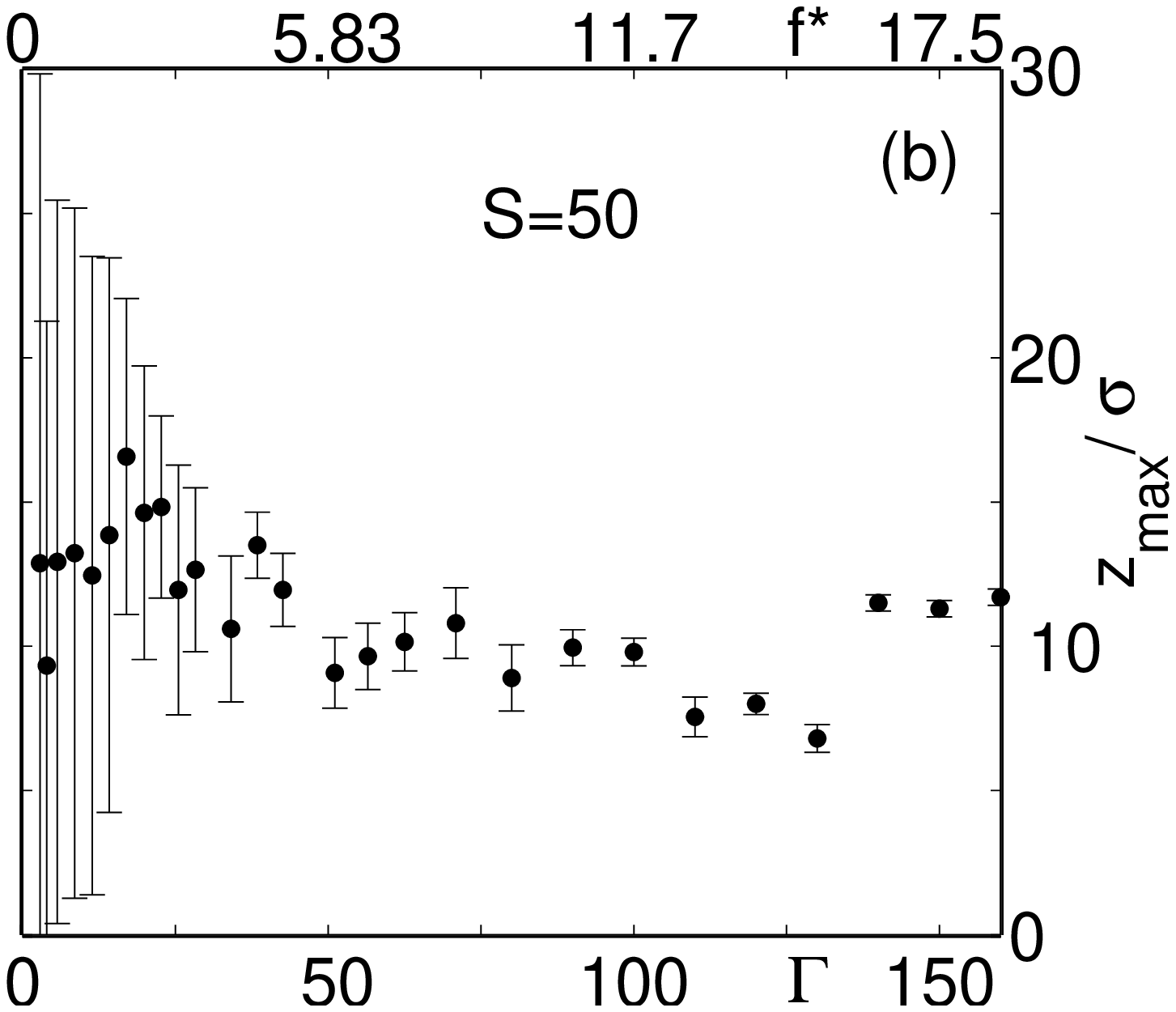}}}\\
\caption{\label{heights} 
Dimensionless height of the maximum  volume fraction $z_{max}/\sigma$  in continuum simulations
as a function of $\Gamma$ (bottom axis) and $f^*$ (top axis), for (a) $S=8$, and (b) $S=50$.  
$z_{max}$ is calculated eight times during a cycle; points show the average of these times
and error bars show standard deviation. 
}
\end{figure}

To examine the effects of this transition on density-inversion, 
we vary $\Gamma$ and plot $z_{max}$ during eight equally spaced times during a cycle in Fig.~\ref{heights}, averaged over 100 cycles.
For both $S=8$ and $S=50$, low-$\Gamma$ shaking produces strong time dependence, as seen by a large standard deviation of
 $z_{max}$. As $\Gamma$ increases, the layer transitions to a steady state in which the height of the maximum does not 
vary significantly with time.  
In both cases, the peak of the layer contacts the plate for low-$\Gamma$ shaking, 
while the peak of the layer is far from the plate for high $\Gamma$.
$S=50$ yields consistently higher $z_{max}$  than does the lower shaking strength $S=8$, indicating 
a more significant density inversion.  
For given $\Gamma$, the amplitude of the plate is much larger
for $S=50$ than for $S=8$, which may account for the larger variations in height
for $S=50$.  The transition also appears to occur at higher $f^*$ for $S=50$, indicating that
the critical frequency above which the layer becomes time-independent differs for differing shaking strengths.

We conclude that for a given shaking strength $S$, time-dependent continuum simulations
exhibit a transition from a strongly time-dependent 
oscillatory state at low $\Gamma$ to a density inverted state that is nearly decoupled from the oscillation
of the plate at high $\Gamma$.
While previous studies only successfully used continuum theory to describe the high-$\Gamma$ state,
we find that time-dependent continuum simulations also describe the low-$\Gamma$ state
and produce results consistent with MD simulations in both regimes.  
The transition to a steady-state density inversion occurs at different driving frequencies
for different shaking strengths; a future systematic study of varying shaking strengths could contribute to our
understanding of this transition.

This work was supported by the Research Corporation for Science Advancement.

\bibliography{paper}

\end{document}